\documentstyle[a4,12pt]{article}

\newcommand{\be}{\begin{equation}}
\newcommand{\ee}{\end{equation}}
\newcommand{\bea}{\begin{eqnarray}}
\newcommand{\eea}{\end{eqnarray}}
\newcommand{\nn}{\nonumber}

\newcommand{\bra}[1]{\left\langle #1 \right|}
\newcommand{\ket}[1]{\left| #1 \right\rangle}
\begin{document}
\vspace*{1cm}
\rightline{BARI-TH/93-152}
\rightline{September 1993}
\vspace*{2cm}
\begin{center}
  \begin{Large}
  \begin{bf}
QCD CALCULATION OF HEAVY MESON \\
\vskip .2cm
SEMILEPTONIC TRANSITIONS \\
  \end{bf}
  \end{Large}
  \vspace{8mm}
  \begin{large}
P. Colangelo \footnote{Talk delivered at the Third Workshop on a Tau-Charm
Factory, Marbella (Spain) 1-6 june 1993.}\\
  \end{large}
  \vspace{6mm}
 Istituto Nazionale di Fisica Nucleare, Sezione di Bari, Italy\\

\end{center}
\begin{quotation}
\vspace*{1.5cm}
\begin{center}
  \begin{bf}
  ABSTRACT
  \end{bf}
\end{center}
\vspace*{0.5cm}

\noindent
I review the QCD sum rules calculations of the form factors governing the
semileptonic decays of charmed and beauty mesons.
In particular, I discuss the predicted
dependence of the various form factors on
$q^2$ and how it can be obtained for the form factors of vector
and axial currents.
In some cases the $q^2$ dependence, computed by QCD sum rules,
 is different from the outcome of lattice QCD.
A Tau-Charm factory could be an efficient tool to study this aspect of the
charmed meson transitions.
\noindent
\end{quotation}
\newpage
\noindent{\Large{\bf{QCD CALCULATION OF HEAVY MESON }}}
\vskip .2cm
\noindent{\Large{\bf{SEMILEPTONIC TRANSITIONS}}}
\vskip 1.0cm
{\it{Pietro Colangelo}}
\vskip 0.0cm
{ INFN, Sezione di Bari, Italy.}
\vskip 1.0cm

\leftskip=2truecm
\rightskip=2truecm

\noindent {\bf Abstract}
\vskip .2cm
\noindent
I review the QCD sum rules calculations of the form factors governing the
semileptonic decays of charmed and beauty mesons.
In particular, I discuss the predicted
dependence of the various form factors on
$q^2$ and how it can be obtained for the form factors of vector
and axial currents.
In some cases the $q^2$ dependence, computed by QCD sum rules,
 is different from the outcome of lattice QCD.
A Tau-Charm factory could be an efficient tool to study this aspect of the
charmed meson transitions.

\leftskip=0truecm
\rightskip=0truecm

\vskip 1.5cm
\pagestyle{empty}

In this review I briefly describe
how QCD sum rules can be used to evaluate the hadronic matrix elements
governing the semileptonic decays of charmed and beauty mesons.
These decays
can be used to determine some elements of the Cabibbo-Kobayashi-Maskawa
matrix, which are fundamental parameters of the Standard Model, and
to check the V-A structure of the weak interactions.
Moreover, they allow us to investigate
the non perturbative structure of the strong dynamics, since the
hadronic matrix elements can be measured and then compared
to the predictions of
the various theoretical approaches used to handle the difficult
problem of describing the confining interactions.

QCD sum rules \cite{SVZ}, as well as lattice QCD
\cite{Martinelli}, are rooted in the QCD
framework of the strong interactions and provide predictions
from first principles; in this respect they differ from
the other approaches to the hadronic dynamics,
e.g. potential models, where the reference to QCD is
diluted in "ad hoc" assumptions.
On the other hand, QCD sum rules are different
 from lattice QCD since they use
analytic methods, whereas lattice QCD heavily relies on numerical
calculations. A general opinion is that
the two approaches  are complementary to each other; as a matter of fact
the predictions of the first one, as we shall see in the following,
cannot be improved to an
arbitrary accuracy, whereas lattice QCD is restricted by the present
computer facilities.

In order to summarize the main results obtained by different QCD sum rules
calculations for the semileptonic transitions
$D \to (K, K^*) \ell \nu$ and
$D \to (\pi, \rho) \ell \nu$ (which mainly concern a Tau-Charm factory)
let us fix the notations. In the
Bauer-Stech-Wirbel parameterization \cite{WSB} the hadronic matrix elements
involved in $D \to (K, K^*) \ell \nu$
 can be written as follows:
\bea
\bra{K(p^\prime)} J_{\mu} \ket{{ D}(p)}\;&=&\;F_1 (q^2) \;
(p+p^\prime)_\mu +
\;{m_{D}^2 - m_K^2 \over q^2} q_{\mu} \; \lbrack F_0 (q^2) - F_1 (q^2)
\rbrack  \label{zero} \\
\bra{K^* (p^\prime )} J_{\mu} \ket{{ D}(p)}\;&=&\;{2 V(q^2)\over m_{D}
+ m_{K^*}} \epsilon_{\mu \alpha \rho \sigma} \epsilon^{\ast \alpha}
p^{\rho}
p^{\prime \sigma} -\nn \\
&-&\;i \Big[ (m_{D} + m_{K^*}) A_1 (q^2) \epsilon_{\mu}^{\ast} - {A_2 (q^2)
\over m_{D} + m_{K^*}} (\epsilon^{\ast}\cdot p) (p + p^\prime)_\mu -\nn \\
&-&\;(\epsilon^{\ast}\cdot p_{D}) {2 m_{K^*} \over q^2} q_\mu
\big(A_3 (q^2) - A_0 (q^2)\big) \Big]  \label{first}
\eea
\noindent
where $q^2 = (p - p^\prime)^2$ and $J_\mu = {\bar s} \gamma_\mu (1 -
\gamma_5) c$ \hskip 5pt ; $\epsilon$ is the $K^*$ meson polarization
 vector.
To avoid unphysical poles at $q^2=0$
the conditions $F_1 (0)\; = \; F_0 (0)$ and
$A_3 (0)\; = \; A_0 (0)$ must be implemented;
$A_3$ can be expressed in terms of $A_1$ and $A_2$:
\be
A_3 (q^2)\;=\;{m_{D}+m_{K^*} \over 2 m_{K^*}} A_1 (q^2) -
{m_{D}-m_{K^*} \over 2 m_{K^*}}
A_2 (q^2) \hskip 5pt . \label{third}\ee
\noindent
In the limit of massless charged leptons the relevant form factors are
$F_1, V, A_1$ and $A_2$. The matrix elements of the transitions
$D \to (\pi, \rho) \ell \nu$ and of the $B$ meson semileptonic decays
can be written as in
Eqs.(\ref{zero},\ref{first}) with obvious changes.

The calculation of the form factors in
(\ref{zero},\ref{first}) has been first performed using potential
models \cite{WSB, KS, ISGW, AW}, \cite{Pene}.
In this approach only few kinematical configurations
can be treated analytically; as a matter of fact, in order
to evaluate a hadronic matrix element as in (\ref{zero},\ref{first})
in a range of $q^2$, typically  one has to compute
an overlap integral involving the wave functions of mesons of arbitrary
momentum.
However, in potential models such wave functions are determined for mesons in
particular kinematical configurations (e.g. for mesons at rest, or in the
infinite momentum frame), and cannot be given for states of
arbitrary momentum since
the problem of performing their relativistic boost is not solved on general
grounds.
 For this reason potential models provide the value of form factors
at the maximum recoil point $q^2=0$ \cite{WSB, KS} or at
the zero recoil point $q^2=q^2_{max} =(m_D - m_{K,K^*})^2$
\cite{ISGW, AW, CNT}, and the functional $q^2$
dependence (polar, multipolar,
exponential) is assumed invoking nearest pole dominance, QCD
counting rules, etc. For example, in the BWS model \cite{WSB}
all the form factors are assumed to have a polar dependence:
\be F_i(q^2)= {F_i(0) \over 1 - q^2/m^2_{pole}} \label{ff} \ee
with the pole given by the nearest resonance in the t-channel \cite{Pene}.
It is worth reminding that the $t-$dependence of the
heavy meson semileptonic form
factors is an important information employed e.g.
in the framework of the
heavy quark effective theory coupled to chiral symmetry, when
 $B\ \to \pi \ell \nu$ and $D\ \to \pi \ell \nu$ are related: in this case an
extrapolation is needed from zero recoil, where predictions can
be derived, to the maximum recoil point where experimental data are available
\cite{Pene,Casal}.

Using three-point QCD sum rules the form factors of the transition
 $D \to (K, K^*) \ell \nu$ at $q^2=0$
 have been first computed in \cite{Aliev, Aliev1} and then in
\cite{Dosh, Dosh1}.
Since the method is general, also the form factors of
 $D \to (\pi, \rho) \ell \nu$ and  of
the semileptonic $B$
decays to negative and positive charmed and non-charmed states have been
computed (at $q^2=0$)
\cite{Dosh, Ovchinnikov, Ovchinnikov1, Slobodenyuk, Narison, Colangelo}.
 In \cite{Paver} the calculation for $D \to K \ell \nu$
has been performed by two
point QCD sum rules.

The results for the $D$ meson transitions are collected in Table 1;
there is an overall agreement among the different estimates (only the central
value of the form factors computed in \cite{Aliev1} are larger than in the
other calculations).
Moreover, the comparison of the results for $D \to K, K^*$ and $D \to \pi,
\rho$ shows that the $SU(3)_F$ breaking effects in the two channels are of the
order of $10-20 \%$.
A comparison with other theoretical approaches and with the
experimental results can be found in these proceedings \cite{Pene}.

As for the $q^2$ dependence,
  already from  the early investigations of the pion electromagnetic
form factor \cite{Radyushkin} QCD sum rules have been proven to be successful
in describing the dependence of hadronic matrix elements on intermediate
(space-like) values of the  transferred momentum  $t$. Also in this respect
QCD sum rules are analogous to  lattice QCD, although this last approach
 has been limited, so far, by statistics and
by small lattice sizes which compel an extrapolation
to $q^2=0$ \cite{Martinelli1}.
In \cite{Dosh1} the $q^2$ dependence has been explicitly studied for the
form factors governing $D \to K^* \ell \nu$, whereas
$B$ decays have been considered in
\cite{Dosh2, Ball}. In \cite{Dosh1, Ruckl} the $q^2$ dependence has also been
studied by light-cone QCD sum rules.

\hskip -1 cm\begin{table}
\centering
\begin{tabular}{|c|c|c|c|c|}
\hline
$F_1^{D \to K}(0)$ & $V^{D \to K^*}(0)$ &
 $A_1^{D \to K^*}(0)$&$ A_2^{D \to K^*}(0)$ &Ref.\\ \hline
$0.6\pm 0.1$&$-$&$-$&$-$ & \cite{Aliev}\\ \hline
$0.8\pm0.2$&$1.6\pm 0.5$&$0.9\pm0.2$&$0.8\pm0.3$& \cite{Aliev1} \\ \hline
$0.6\pm 0.1$&$-$&$-$&$-$ & \cite{Dosh}\\ \hline
$0.6^{+0.15}_{-0.10}$&$1.1\pm 0.25$&$0.5\pm0.15$&$0.6\pm0.15$&
\cite{Dosh1}\\ \hline \hline
$F_1^{D \to \pi}(0)$ & $V^{D \to \rho}(0)$ &
$A_1^{D \to \rho}(0)$&$ A_2^{D \to \rho}(0)$ &Ref.\\ \hline
$0.7\pm 0.2$&$-$&$-$&$-$ & \cite{Aliev1}\\ \hline
$0.75\pm 0.05$&$-$&$-$&$-$ & \cite{Paver}\\ \hline
$0.5\pm0.1$&$1.0\pm 0.2$&$0.5\pm0.2$&$0.4\pm0.1$& \cite{Ball} \\ \hline
\end{tabular}
\caption{\it{QCD sum rules estimates of semileptonic form factors
for $D \rightarrow K, K^*$ and
for $D \rightarrow \pi, \rho$.}} \label{tab:final}
\end{table}
The result of these investigations is that the form factors
$F_1$ and $V$ of the vector
current in $D \to K, K^*$, $D \to \pi, \rho$
 have a polar $t-$dependence as in Eq.(\ref{ff}),
with a pole mass in some agreement with the mass of the first
resonance in the $t$ channel.
The fitted masses are:
$m_{pole}=1.81 \pm 0.10 \; GeV$ for $D \to K$ and
$m_{pole}=1.95 \pm 0.10 \; GeV$ for $D \to K^*$, to be compared to the measured
mass of $D^*_s$: $m_{D^*_s}=2.11 \; GeV$;
$m_{pole}=1.95 \pm 0.10 \; GeV$ for $D \to \pi$ and
$m_{pole}=2.5 \pm 0.2 \; GeV$ for $D \to \rho$, to be compared to
$m_{D^*}=2.01 \; GeV$.
Also the form factors of the vector current in
$B \to \pi, \rho$ transitions have such behaviour; in this case
$m_{pole}=5.25 \pm 0.10 \; GeV$ for $B \to \pi$ and
$m_{pole}=6.6 \pm 0.6 \; GeV$ for $B \to \rho$, to be compared to
$m_{B^*}=5.33 \; GeV$ \cite{Dosh1, Ball}.
This is in agreement with the usual assumption made in BWS and in other
potential models, and with the outcome of lattice QCD \cite{Martinelli1}.

For the form factors $A_1$ and $A_2$ of the axial current in
$D \to K^*, \rho$ and in $B \to \rho$ transitions, QCD sum rules show
the absence of the polar dependence; the form factors are nearly independent
of $q^2$, at odds with the outcome e.g. of lattice simulations
\cite{Martinelli1}. A confirmation of this result comes from an analysis of the
scaling properties of the
$B \to \pi, \rho, K^*$ form factors in the limit $m_b \to \infty$
\cite{Colangelo2}.

This result requires a careful
investigation; from the experimental point of view, there is evidence that
$F_1^{D \to K}$ is polar \cite{e653}, whereas
no information is available
on the $t$ dependence of the form factors of
$D \to K^*$. A description of the
difficulties of such analysis and of the potentialities of a Tau-Charm factory
can be found in these proceedings \cite{Pene, Roudeau}.
{}From the theoretical point of view this behaviour has not found an
explanation, yet.

An interesting problem is to investigate if such anomalous $t$-dependence is
common to all the matrix elements of the axial current
or if it is peculiar of the $0^- \to 1^-$ transitions. To study this problem,
and to show in detail how QCD sum rules can be used to evaluate semileptonic
form factors  I consider the decays:
\bea D^0& \to & \pi^- \; \ell^+ \; \nu_\ell \label{eq1}\\
     D^0 & \to & a_0^- \; \ell^+ \; \nu_\ell \label{eq2} \eea

\noindent where $a_0$ is the $J^P=0^+$ orbital excitation of the pion system
which can be identified with $a_0(980)$.
The decay (\ref{eq1}) is induced by a vector weak current and
(\ref{eq2}) by an axial current
\footnote{The following analysis of $D \to \pi \ell \nu$
is similar to \cite{Aliev, Dosh1, Ball}; the results for
$D \to a_0 \ell \nu$ are new.}.
In terms of form factors the hadronic matrix
elements of (\ref{eq1}) and (\ref{eq2}), keeping only the terms that contribute
for massless charged leptons,
can be written as follows:

\bea <\pi(p^\prime) | V_\mu | D(p)> & = &
 F_1^{D \to \pi} (q^2) \; (p + p^\prime)_\mu + \dots  \label{eq3}\\
i <a_0(p^\prime) | A_\mu | D(p)> & = &
 F_1^{D \to a_0} (q^2) \; (p + p^\prime)_\mu + \dots \; .\label{eq4} \eea

The starting point to compute $F_1^{D \to \pi} (q^2)$ is the three-point
function correlator
\be T^{D \to \pi}_{\mu\nu} (p, p^\prime, q) = i^2 \int dx \; dy \;
e^{i ( p^\prime \cdot x - p \cdot y)} \;
<0| T \{ j_\nu(x) V_\mu(0) j^\dagger_5(y) \} |0> \label{eq5} \ee

\noindent where $j_\nu$ and $j_5$ are local currents of quark fields
with the same quantum numbers of a pion and of a $D$ meson:
$j_\nu(x) = \bar u(x) \gamma_\mu \gamma_5 d(x)$,
$j_5(y) = \bar u(y) i \gamma_5 c(y)$; $V_\mu$ is the vector current
inducing the transition (\ref{eq1}) :
$V_\mu(0) =\bar d(0) \gamma_\mu c(0)$. In analogous way the calculation of
$F_1^{D \to a_0} (q^2)$ starts from:
\be T^{D \to a_0}_\mu (p, p^\prime, q) = i^2 \int dx \; dy \;
e^{i ( p^\prime \cdot x - p \cdot y)} \;
<0| T \{ j_s(x) A_\mu(0) j^\dagger_5(y) \} |0> \label{eq6} \ee

\noindent where $j_s(y) = \bar u(y) d(y)$ and
$A_\mu(0) =\bar d(0) \gamma_\mu \gamma_5 c(0)$
is the axial current which  induces the transition (\ref{eq2}).
It is immediate to generalize the method by changing the
currents in order to compute  different charmed or beauty mesons decays
to vector or axial states.

After a decomposition in Lorentz invariant structures:
\be T^{D \to \pi}_{\mu\nu} (p, p^\prime, q) =
i \; T^{D \to \pi} (p^2, p^{\prime 2}, q^2) \;
(p+ p^\prime)_\mu \; p^\prime_\nu + ....
\label{eq7} \ee

\be T^{D \to a_0}_\mu (p, p^\prime, q) =
i \; T^{D \to a_0} (p^2, p^{\prime 2}, q^2) \; (p+ p^\prime)_\mu  + ....
\label{eq8} \ee
\noindent
the strategy of QCD sum rules is to evaluate
$ T^{D \to \pi} (p^2, p^{\prime 2}, q^2)$  and
$T^{D \to a_0} (p^2, p^{\prime 2}, q^2)$ in two independent ways.
First, taking into account the  analytical properties of
$ T^{D \to \pi} (p^2, p^{\prime 2}, q^2)$ and
$T^{D \to a_0} (p^2, p^{\prime 2}, q^2)$
in the variables
$p^2$ and $p^{\prime 2}$,
a double dispersion relation is written:

\be T (p, p^\prime, q) = {1 \over (2 \pi)^2} \int ds \; ds^\prime
{\rho(s, s^\prime, q^2) \over (s - p^2) (s^\prime - p^{\prime 2})}
\; + \; subtractions .\label{eq9}\ee

\noindent with the spectral function $\rho$ getting contributions from
hadronic intermediate states; in terms of the lowest lying resonances
the physical spectral functions read:

\be  \rho^{D \to \pi}(s, s^\prime, q^2) = (2 \pi)^2 f_\pi f_D {m^2_D \over m_c}
 F_1^{D \to \pi}(q^2) \delta(s-m^2_D) \delta(s^\prime - m^2_\pi)
+ \rho^{D \to \pi}_{cont}(s, s^\prime, q^2) \label{eq10} \ee

\be  \rho^{D \to a_0}(s, s^\prime, q^2) = (2 \pi)^2 f_{a_0} f_D
{m^2_D \over m_c}
 F_1^{D \to a_0}(q^2) \delta(s-m^2_D) \delta(s^\prime - m^2_{a_0})
+ \rho^{D \to a_0}_{cont}(s, s^\prime, q^2) \label{eq11} \ee
\noindent where
$f_\pi=132 \; MeV$ and $f_{a_0}$ is defined by the matrix element
$f_{a_0}=<0|\bar d u | a_0>$.

On the other hand, the correlators (\ref{eq5},\ref{eq6}) are computed in QCD
by an operator product expansion (OPE) at $p^2, p^{\prime 2}, q^2$
spacelike and large:
in this expansion not only the perturbative term is taken into account,
but also higher order corrections given in terms
of vacuum expectation values of quark and gluon fields (condensates) ordered by
dimension and divided  by powers of $p^2$ and  $p^{\prime 2}$.
The result for Eq.(\ref{eq7}), neglecting the light quark masses,
is:
\bea T_{QCD}^{D \to \pi} (p^2, p^{\prime 2}, q^2)
& = & {1 \over (2 \pi)^2} \int ds \; ds^\prime
{\rho^{D \to \pi}_{QCD}(s, s^\prime, q^2) \over (s - p^2)
(s^\prime - p^{\prime 2})}
\; \nonumber \\ \nonumber \\
& - & {<\bar q q> \over 2} [ {1 \over r r^\prime} - {m_0^2 \over 6} \;
({3 m_c^2  \over r^3 r^\prime} +
{2 \over r ^2 r^\prime} - {2 q^2  \over r^2 r^{\prime 2} } )] \label{eq12}\eea
\noindent
where $r=p^2 -m_c^2$, $r^\prime=p^{\prime 2}$.
The spectral integral represents the perturbative contribution to the OPE
 at the lowest order in $\alpha_s$:
\bea  \rho^{D \to \pi}_{QCD}  (s, s^\prime, q^2) & = &
{3 m_c \over 2 \lambda^{3/2} } \Big[ 2 \Delta (u - s^\prime) +
s^\prime (u - 4 s)
\nonumber \\ \nonumber \\
& - & {2 m_c \over \lambda }
\Big(  \Delta^2 (u^2 - 3 u s^\prime + 2 s s^\prime)
\nonumber \\ \nonumber \\
& + & 2 \Delta s^\prime (u^2 - 3 u s + 2 s s^\prime)
+ 3 s s^{\prime 2} ( 2 s -u) \Big) \Big] \label{eq13} \eea
\noindent ($\Delta=s - m_c^2$, $u=s +s^\prime - q^2$ and
$\lambda= u^2 - 4 s s^\prime$); in principle,
higher order $\alpha_s$ corrections can also be
included, although their explicit calculation is not available, yet.
The terms proportional to $<\bar q q> $ and to
$m_0^2 <\bar q q> $ (using the notation
$ <\bar q g \sigma G q> = m_0^2 <\bar q q> $) are the first two power
corrections (in terms of operators of dimension $D=3$ and $D=5$, respectively)
which parameterize the deviations from the asymptotically free
behaviour. In QCD sum rules these condensates are universal parameters:
they are independent of the channel, can be fixed
from low energy phenomenology and compared to the evaluation by
lattice QCD. Their value is known only for low dimensional
operators, whereas the higher dimensional condensates are generally
estimated by using factorization.
In practical cases the number of the power corrections
that can be included in the expansion is  limited;
for example, in Eq.(\ref{eq13}) the $D=6$ term is known and gives
a negligible contribution, whereas the contribution of $D=4$ operator has been
estimated for the Borel transformed sum rule, only \cite{Ball}.
The truncation in the series of the non-perturbative corrections is a
limitation of the QCD sum rules approach.
The heavy quark masses are parameters to be fixed;
we use $m_c=1.35 \; GeV$ and
$m_b=4.6 \; GeV$
\footnote{The dependence of the QCD sum rules predictions for the leptonic
constants on the heavy quark masses is discussed in
\cite{Colangelo1}.}.

The result for Eq.(\ref{eq8}) is obtained in analogous way:
\newpage
\bea T_{QCD}^{D \to a_0}  (p^2, p^{\prime 2}, q^2) & = &
{1 \over (2 \pi)^2} \int ds \; ds^\prime
{\rho^{D \to a_0}_{QCD}(s, s^\prime, q^2) \over (s - p^2)
(s^\prime - p^{\prime 2})} \;
\nonumber \\ \nonumber \\
& - & {m_c <\bar q q> \over 2} [ {1 \over r r^\prime} - {m_0^2 \over 6} \;
({3 m_c^2  \over r^3 r^\prime} +
{4 \over r^2 r^\prime} +
{2 \over r r^{\prime 2}}
\nonumber \\ \nonumber \\
& + & {2 (m_c^2 - q^2)  \over r^2 r^{\prime 2} } ) ]
\label{eq14}\eea
\noindent
with the spectral function:
\be \rho^{D \to a_0}_{QCD}(s, s^\prime, q^2)  =
{s^\prime  \over \lambda^{3/2} } \Big[ m_c^2 (s - s^\prime) -
q^2 (2 s - m_c^2)   \Big] \hskip 5pt . \label{eq15} \ee

Invoking duality, it is assumed
 that $\rho_{cont}$ in (\ref{eq10},\ref{eq11}),
which includes the contribution of the higher resonances and of the
continuum of states, is equal
to $\rho_{QCD}$ given in (\ref{eq13},\ref{eq15})
for all values of $s, s^\prime, q^2$ but for a region where the lowest lying
resonances dominate; a model for this region is:

\bea m_c^2 < & s & < s_0 \nonumber \\
0 < & s^\prime  & <
min \big( s^\prime_0 , \; {(s-m_c^2) (m_c^2 - q^2) \over m_c^2} \big)
\hskip 5pt. \label{eq16} \eea

\noindent $ s_0$ and $ s^{\prime}_0$ are thresholds which separate the
domain of the resonance from continuum; their exact position is not
known, although indications can be derived from the experimental or
theoretical spectrum in a given channel.
For example, for $D \to \pi$ it can be assumed
 that $s_0$ is not larger than the
mass squared of the first resonance above $D$ coupled to the pseudoscalar
current, $s_0=6-7 \; GeV^2$ and that $s_0^\prime$ is around the
$\rho, \omega$ mass squared,
$s_0^\prime = 0.6 - 0.7 \; GeV^2$.
For $a_0$ we choose:
$s_0^\prime = 1.6 - 1.7 \; GeV^2$, and for $B$:
$s_0 = 33 - 36 \; GeV^2$.

Assuming duality, a rule is obtained where the form factors in
(\ref{eq10},\ref{eq11}) are given in terms of the QCD quantities
in (\ref{eq12}-\ref{eq14})
(quark masses, condensates, eventually $\alpha_s$)
and of the leptonic constants $f_\pi, f_D, f_{a_0}$.
The subtraction terms
(polynomials in $p^2$ or $p^{\prime 2}$),
which can be  present in (\ref{eq10},\ref{eq11}),
are removed by performing a double Borel transform in the
variables $-p^2$, $-p^{\prime 2}$:
\be
{\cal B} = {(-p^2)^n \over (n-1)!} \Big( {d \over d p^2} \Big)^n
{(-p^{\prime 2})^m \over (m-1)!} \Big( {d \over d p^{\prime 2}} \Big)^m
\label{eq17}
\ee
\noindent in the limit
$-p^2, -p^{\prime 2}  \to \infty$, $n, m \to \infty$,
keeping ${-p^2 / n}=M^2$ and
${-p^{\prime 2} / m}=M^{\prime 2}$ fixed.
This transformation has also the property of
factorially suppressing the higher order
power corrections in the operator product
expansion,  and of  enhancing the contribution of the low lying states in the
perturbative term.
The Borel transformed sum rules for
$F_1^{D \to \pi}$  and $F_1^{D \to a_0}$  read:
\bea
f_\pi f_D {m^2_D \over m_c}  F_1^{D \to \pi}(q^2)
e^{- { m^2_D \over M^2} - { m^2_\pi \over M^{\prime 2}} } & = &
  {1 \over (2 \pi)^2} \int_D ds \; ds^\prime
\rho^{D \to \pi}(s, s^\prime, q^2) \;
 e^{- {s \over M^2} - { s^\prime \over M^{\prime 2}} } \nonumber \\
& - & {<\bar q q> \over 2} \; e^{ - {m^2_c \over M^2} } \;
\big[ 1 - {m_0^2 \over 6} \; ({3 m_c^2  \over 2 M^4 } -
{2 \over M^2} - {2 q^2  \over M^2 M^{\prime 2} } ) \big] \nonumber \\
\label{eq18} \eea
\noindent and
\bea
 f_{a_0} f_D {m^2_D \over m_b} F_+^{D \to a_0}(q^2)
e^{- { m^2_D \over M^2} - { m^2_{a_0} \over M^{\prime 2}} } & = &
{1 \over (2 \pi)^2} \int_D ds \; ds^\prime
\rho^{D \to a_0}(s, s^\prime, q^2) \;
 e^{- {s \over M^2} - { s^\prime \over M^{\prime 2}} } \nonumber \\
 & - & {<\bar q q> \over 2} \; e^{ - { m^2_c \over M^2}} \;
\big[1- {m_0^2 \over 6} ({3 m_b^2  \over 2 M^4 } - {4 \over M^2} -
\nonumber \\
& - &
{2 \over M^{\prime 2}} + {2 ( m_c^2 - q^2)  \over M^2 M^{\prime 2} } )\big]
\hskip 5pt .\label{eq19}\eea
\noindent

\begin{figure}[t]   
    \begin{center} \setlength{\unitlength}{1truecm} \begin{picture}(6.0,6.0)
\put(-6.0,-7.5){\special{dpi.ps hoffset=36 hscale=100 vscale=100}}
       \end{picture} \end{center} 	\vskip 4.cm \caption[]
{\it {$q^2$ dependence of the form factors $F_1^{D  \to \pi}$
(continuous line) and $F_1^{D \to a_0}$ (dashed line).}}
\protect\label{dpi} \end{figure}

The last quantities to be fixed are the leptonic constants.  Some of them,
 as $f_\pi$, come from
experiment;  others, as $f_D$,  can be obtained from
two-point QCD sum rules \cite{Colangelo1}.
The analysis of Eqs.(\ref{eq18},\ref{eq19}) proceeds by looking
for a range of parameters $M^2, M^{\prime 2}$ (duality window)
where the values of the form
factors do not depend on the Borel parameters
and on the thresholds $s_0, s^\prime_0$; in this window a
hierarchy between the perturbative $D=0$ and the non perturbative $D=3, D=5$
terms should be verified in order to be confident on the convergence
of the series of the power corrections; moreover,
the perturbative integral should be larger than the contribution of the
continuum.
Other conditions can be imposed to restrict the dependence
of the predictions on the
the parameters of the method; for example, one could check that the
sum rule for the mass of the resonances, obtained by taking logarithmic
derivatives with respect to the Borel parameters, gives a result in agreement
with the experimental mass.

If all these requirements are
fulfilled a prediction for the form factors in the deep Euclidean region
$q^2 \le 0$
is obtained. Moreover, as discussed in \cite{Dosh}, the analysis
of the sum rule
can also be done for positive values of $t$, since long distance effects in the
$t-$channel can only be relevant near the threshold
$t_{th}\simeq m_c^2$ (for $D \to K, K^*, \pi )$ or
$t_{th}\simeq m_b^2$ (for $B \to \pi, \rho)$. In this way indications can be
obtained about the
$q^2$ dependence of the form factors in a
quite large range of $q^2$.

\begin{figure}[t]   
    \begin{center} \setlength{\unitlength}{1truecm} \begin{picture}(6.0,6.0)
\put(-6.0,-7.5){\special{bpi.ps hoffset=36 hscale=100 vscale=100}}
       \end{picture} \end{center} 	\vskip 4.cm \caption[]
{\it {$q^2$ dependence of the form factors  $F_1^{B  \to \pi}$
(continuous line) and $F_1^{B \to a_0}$ (dashed line).}}
\protect\label{bpi} \end{figure}

The $t$ dependence of
$F_1^{D \to \pi}$ and $F_1^{D \to a_0}$ is depicted in fig.\ref{dpi}.
As already observed in \cite{Dosh1,Ball}
$F_1^{D \to \pi}$ displays a $t$ dependence that can be fitted
with a pole. The interesting point, here, is
that also $F_1^{D \to a_0}$, which is related to an axial current,
 displays a polar behaviour with a fitted pole mass
$m_{pole}\simeq 1.9 \; GeV$; however, this mass is smaller than the mass of
$D^{**}(1^+)$ which is the first resonance in the $t$ channel:
$m_{D^{**}(1^+)}=2.42 \; GeV$.
A similar result is obtained for $B \to \pi$ and $B \to a_0$
(fig.\ref{bpi}); in the second
case the fitted pole mass is
$m_{pole}\simeq 6.2 \; GeV$.

The conclusion is that there is no common behaviour for the form factors of the
axial current (and, presumably, of the vector current); the $q^2$ dependence of
the various form factors must be computed, since general arguments of
dominance, universality, etc., could fail.

The experimental investigations on semileptonic $D-$meson decays are in their
initial status. The predictions made by QCD sum rules, or by other theoretical
approaches, have to be checked not only because of their own interest, but also
in the light of their implications for the phenomenology of $B$ and other
heavy mesons. A Tau-Charm factory, with its potentialities in collecting
large samples of decaying charmed mesons, could be an ideal tool for such
investigations.

\newpage
\noindent {\bf Acknowledgments}

\noindent It is a pleasure to thank C.A.Dominguez, G.Nardulli and N.Paver
for enlightening discussions. Thanks are also due to P.Blasi and F.De Fazio
for their collaboration on the topics discussed here.

\end{document}